\documentclass{article}
\usepackage[a4paper, total={6in, 9in}]{geometry}
\usepackage[utf8]{inputenc}
\usepackage[usenames,dvipsnames]{color,xcolor}
\usepackage{newclude}
\usepackage{booktabs}
\usepackage{url}
\usepackage[normalem]{ulem}
\usepackage{hyperref}
\hypersetup{colorlinks,linkcolor={MidnightBlue},citecolor={Mahogany},urlcolor={MidnightBlue} } 
\usepackage{graphicx}

\usepackage[font=small,labelfont=bf]{caption}

\usepackage{array}

\usepackage{graphicx}

\usepackage{lscape}

\usepackage{bibunits}
\begin{document}

\newlength{\drop}

  \begin{titlepage}
    \drop=0.2\textheight
    \centering
    \vspace*{\baselineskip}
    \vspace*{100pt}
    \rule{\textwidth}{1.6pt}\vspace*{-\baselineskip}\vspace*{2pt}
    \rule{\textwidth}{0.4pt}\\[\baselineskip]
      
    \LARGE The interplay between topography and contact line pinning mechanisms on flat and superhydrophobic surfaces
    \rule{\textwidth}{0.4pt}\vspace*{-\baselineskip}\vspace{3.2pt}
    \rule{\textwidth}{1.6pt}\\[\baselineskip]

    \vspace*{1\baselineskip}
    \large Mahya Meyari$^{1*}$, Camelia Dunare$^{2}$, Khellil Sefiane$^{3}$, Simon Titmuss$^{1}$, Job H. J. Thijssen$^{1*}$
    \\  \vspace*{1\baselineskip}
    \small{$^{1}$  SUPA School of Physics and Astronomy, The University of Edinburgh, Edinburgh EH9 3FD, UK \\}
    $^{2}$  Scottish Microelectronics Centre, School of Engineering, The University of Edinburgh,  Edinburgh EH9 3FF, UK\\
    $^{3}$  Institute for Multiscale Thermofluids, School of Engineering, The University of Edinburgh, Edinburgh EH9 3FD, UK\\
    
    \textbf{$^{*}$E-mail:  }\href{mailto:mahya.m@ed.ac.uk}{mahya.m@ed.ac.uk} 
   \textbf{|}  \href{mailto:j.h.j.thijssen@ed.ac.uk}{j.h.j.thijssen@ed.ac.uk}

    \vspace*{2\baselineskip}
   \normalsize{ {\Large{ \textbf{Abstract}}} \\
   Wettability of a surface depends on both surface chemistry and topography. To move a three-phase contact line, a de-pinning force needs to be applied, which is of practical importance in various applications. However, a unified understanding and description of the de-pinning force on both flat and superhydrophobic surfaces is still lacking. This study aims to bridge the existing gap in our understanding of the three-phase contact line pinning on  flat and microstructured superhydrophobic surfaces. The findings indicate that a general model, based on two different pinning mechanisms, can describe the pinning force on both flat and microstructured surfaces. We compare the general model against experimental data from literature, as well as our experiments on flat and microstructured surfaces coated with a liquid-like layer of grafted polymer chains.
   While this theoretical framework can be useful for designing micro-engineered surfaces on which the contact line behaviour is important, it also provides a potential experimental strategy to distinguish the contribution of defects from that of molecular re-orientation to contact line pinning on a given solid material.}
\vspace*{1\baselineskip}

    \vfill
    
    {\scshape \Large October 2024} \\
    {\large}\par
\end{titlepage}

\begin{center}

\end{center}

\begin{bibunit}[IEEEtran]

\section{Introduction}
\begin{figure}[h]
\begin{center}
\includegraphics[width=0.6\linewidth]{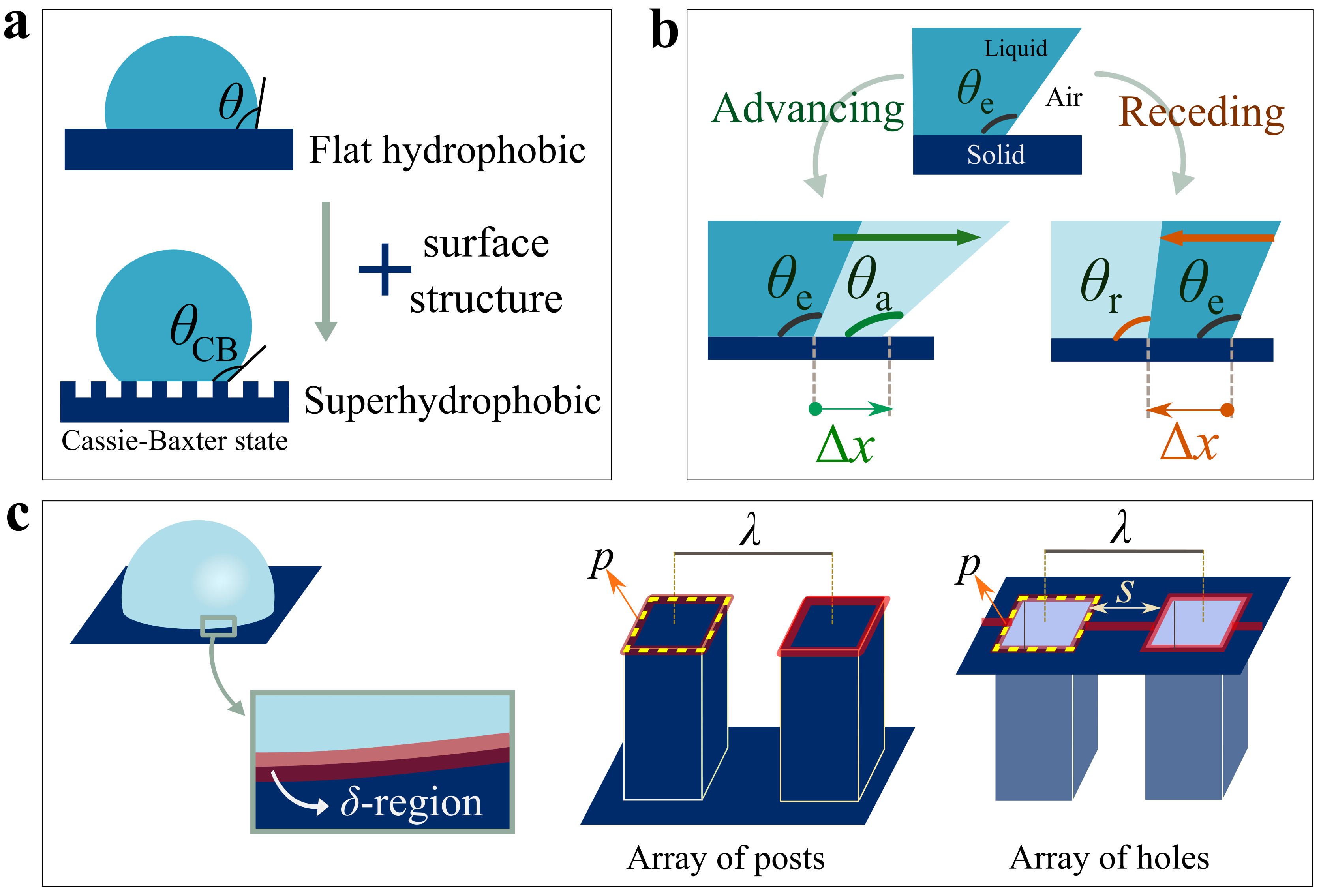}
\caption{\textbf{a)} Adding surface structure to a chemically hydrophobic surface and making it superhydrophobic with the droplet sitting in the suspended (or Cassie-Baxter) state; \textbf{b)} green and orange arrows show the de-pinning force applied to the advancing and receding contact lines, respectively; \textbf{c)} formation of the $\delta$-region on a flat surface, microposts and microholes. $\delta$-region is a narrow region right under the microscopic contact line and its normalised length $\delta$, is the total length of the three-phase contact line per unit length of the macroscopic contact line.}
    \label{fig_delta}
\end{center}
\end{figure}

Surface wettability is primarily quantified by the contact angle between the solid substrate and the liquid-air interface. Although on a flat surface the contact angle is strictly limited by the surface chemistry, adding topographical features or structures to the chemically-hydrophobic surface, as shown in Figure \ref{fig_delta}a, can allow for the observation of contact angles larger than the chemistry-dictated limit — making the surface \textit{superhydrophobic} \cite{Bonn2009WettingSpreading}. It is well known that on real, non-ideal surfaces, whether flat or structured, the contact angle does not have a unique value, but falls within a range called contact angle hysteresis, which is limited by the advancing and receding contact angles. Contact angle hysteresis is the reason for contact line pinning, and understanding how it changes with surface topography is important for various applications such as water harvesting \cite{Seo2014InfluenceDewing}. Energy minimisation and balancing the effective forces are the two approaches used in the literature to deal with the concept of contact line pinning \cite{Bonn2009WettingSpreading, Shanahan1995SimpleHysteresis, Butt2022ContactHysteresis, Long2006OnHysteresis}. The hysteresis force, or what we call the (de-)pinning force, can be described as the force (per unit length) required to de-pin a receding and an advancing contact line \cite{Bonn2009WettingSpreading}:
\begin{eqnarray} \label{Fd}
F_{d}=F_{d}^{rec.} + F_{d}^{adv.} = \gamma(\cos \theta_{r}-\cos \theta_{e}) +\gamma(\cos \theta_{e}-\cos \theta_{a}) = \gamma(\cos \theta_{r}-\cos \theta_{a})
\end{eqnarray}
where $\gamma$ is the liquid surface tension, $\theta_{e}$ stands for the equilibrium contact angle, and $\theta_{a}$ and $\theta_{r}$ denote the advancing and receding contact angles, respectively.  
$F_{d}$ as expressed above is the macroscopic representation of the microscopic pinning events that act on unit lengths of the advancing and receding contact lines. We note that $F_{d}$ as defined in Equation (\ref{Fd}) is not a force to be measured directly. Nonetheless, it can be calculated using the measured values of the advancing and receding contact angles. $F_{d}$ characterises the overall pinning strength without having to know the value of the equilibrium contact angle to which we do not have access. The above relation can also be viewed as the normalised retention force of a two-dimensional droplet. A schematic illustration of the contact lines de-pinning is given in Figure \ref{fig_delta}b.

\begin{figure}[h]
\begin{center}
\includegraphics[width=1\linewidth]{ 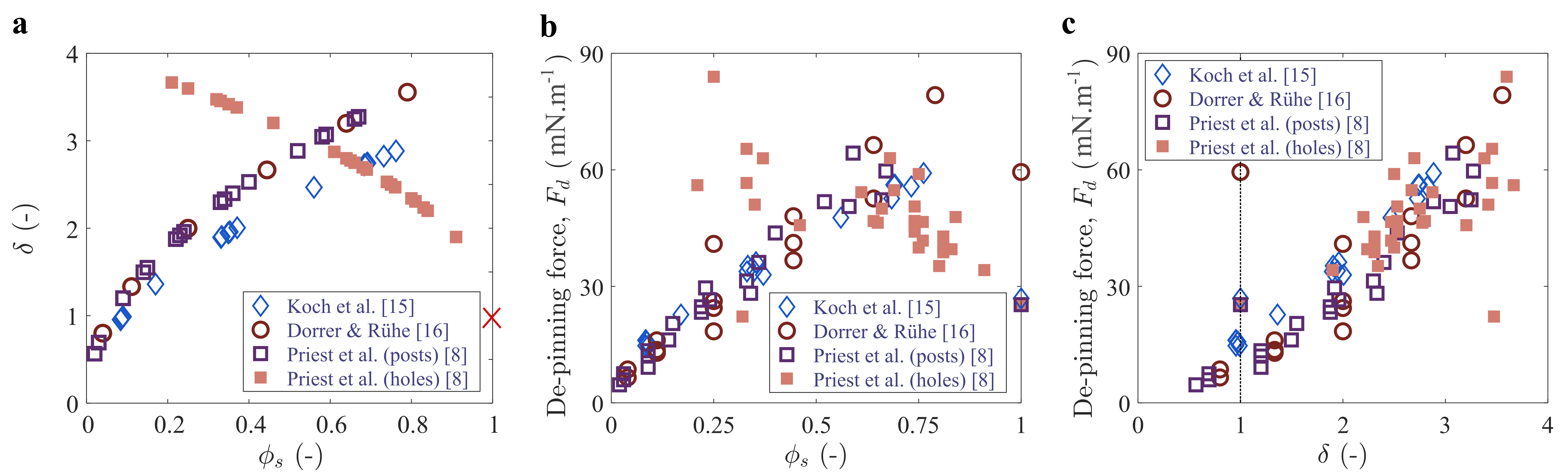}
\caption{Change of topography and $F_{d}$ on flat and microstructured surfaces: \textbf{a)} the relation between $\delta$ and $\phi_{s}$ on four sets of microstructured surfaces. By excluding the flat surface, each of the parameters $\phi_{s}$ and $\delta$ can be defined in terms of another. \textbf{b)} Change of $F_{d}$ with solid surface fraction ($\phi_{s}$) and \textbf{c)} change of $F_{d}$ with the normalised length of the maximal three-phase contact line ($\delta$). The values of the $F_{d}$ are calculated according to Equation (\ref{Fd}) and based on the geometries and the values of $\theta_{a}$ and $\theta_{r}$ reported in references \cite{Priest2009AsymmetricSurfaces,Koch2014ModelingSurfaces, Dorrer2006AdvancingSurfaces}.}
    \label{fig_litData}
\end{center}
\end{figure}

The observed contact angle between a structured surface and a droplet suspended across its features (such as the one illustrated in Figure \ref{fig_delta}a) is usually predicted by the famous Cassie-Baxter equation \cite{Cassie1944WettabilitySurfaces, Lafuma2003SuperhydrophobicStates}, which involves the fraction of the solid-liquid interfacial contact — the solid surface fraction, $\phi_{s}$. Consequently, the de-pinning force has sometimes been considered to be a function of the solid surface fraction on different structured surfaces \cite{Priest2009AsymmetricSurfaces,Cansoy2011EffectSurfaces,Qiao2017FrictionSurfaces}. However, Xu and Choi showed that solid surface fraction ($\phi_{s}$) cannot be used to relate the de-pinning force on a flat surface (i.e. $\phi_{s}=1$) to the structured ones \cite{Xu2012FromSurfaces}. Instead, they employed another parameter, the length of the maximal three-phase contact line per unit length of the macroscopic contact line, $\delta$ \cite{Extrand2002ModelSurfaces}. Xu and Choi used  $\delta$ to describe the de-pinning force on flat and microstructured surfaces which determines the slipperiness or stickiness of a surface \cite{Xu2012FromSurfaces}. They demonstrated, experimentally, that a static contact line on a superhydrophobic surface is initially pinned to the perimeter of the features \cite{Xu2012FromSurfaces}. On an isotropic array of microstructures (either posts or holes), $\delta$ is reduced to:
\begin{equation} \label{eq_delta}
\delta= \frac{L_{TCL}}{L} = \frac{n(p+s)}{L} = \frac{p+s}{\lambda} 
\end{equation}
where $L_{TCL}$ denotes the total length of the microscopic three-phase contact line around the macroscopic contact line, and $L$ is the length of the macroscopic contact line. \textit{n} and \textit{p} are the number of microstructures under the macroscopic contact line and the top perimeter of each structure, respectively. $\lambda$ stands for the pitch between microstructures, and the length of the solid connection between two successive microstructures is denoted by \textit{s} (i.e. \textit{s} = 0 on an array of posts). Hence, on a flat surface $\delta$ equals 1. In this paper we consider $\delta$ as the normalised length of a \textit{region} with a very small width $w_{\delta}\ll 1 \; \mu$m, and call that $\delta$-region hereafter. $\delta$-region is schematically illustrated in Figure \ref{fig_delta}c along with the parameters of Equation (\ref{eq_delta}). While it was suggested that the de-pinning force (or stickiness) on the surfaces are proportional to $\delta$  \cite{Xu2012FromSurfaces,Sarshar2019DepinningModels}, several counter-examples can be found in the literature where $\delta$ fails to describe the stickiness and de-pinning force on both flat and superhydrophobic surfaces \cite{Priest2009AsymmetricSurfaces, Qiao2017FrictionSurfaces, Zhao2012Effectrobustness, Koch2014ModelingSurfaces, Dorrer2006AdvancingSurfaces}. Figure \ref{fig_litData} shows some of the reported experimental results that are not always proportional to $\delta$ nor to $\phi_{s}$.

 Despite the insightful studies on contact angle hysteresis and pinning on structured surfaces \cite{Sarshar2019DepinningModels, Jiang2019GeneralizedSurfaces}, to the best of our knowledge, a unified understanding of the de-pinning forces on flat and superhydrophobic surfaces is lacking. Focusing on the liquids in the suspended (i.e. Cassie-Baxter) state, we combine the defect (or blemish) model \cite{deGennes2004CapillarityPhenomena,Joanny1984AHysteresis} and the molecular re-orientation \cite{Chen1991MolecularSurfaces} or adaptation theory \cite{Butt2018AdaptiveWetting} to explain and unify the de-pinning force on both flat and microstructured surfaces. The model is supported by the measurements taken on micropillared surfaces of varying geometries covered with a Quasi-Liquid Surface (QLS) coating and the data available in the literature. The results suggest that the de-pinning force can be considered as a function of both $\phi_{s}$ and $\delta$, and that function can describe the force on flat surfaces, arrays of posts, and arrays of holes. 

\section{Model for the de-pinning force}
Considering that pinning results from the existence or formation of some regions with a different wettability, the total energy that must be expended to displace the contact line by $\Delta x$ is a function of the density of those regions and their pinning strengths. This function can be expressed in the form of a weighted summation of the energy dissipating events, as in Equation (\ref{general}). We note that as it is the pinning regions in the vicinity of the macroscopic contact line that determine the pinning force, we can write: 

\begin{equation} \label{general}
F_{d} L \Delta x =  \sum d_{i} I_{i} L \Delta x
\end{equation}
where \textit{i} indicates the type of the pinning region and $d_{i}$ is its relevant density around the macroscopic contact line. $I_{i}$ stands for the strength of $i$ (i.e. the amount of energy required for the contact line to pass each unit \textit{i}). To derive a general model describing the pinning force on different surfaces, we go back to defects or blemishes \cite{deGennes2004CapillarityPhenomena, Joanny1984AHysteresis} and molecular re-orientation or adaptation \cite{Butt2018AdaptiveWetting, Tadmor2009MeasurementSubstrate, Tadmor2021OpenForces} as different reasons for pinning on both flat and microstructured surfaces.

\subsection {Blemishes on the solid, $F_{B}$}
According to Joanny and de Gennes \cite{Joanny1984AHysteresis}, contact line pinning is caused by defects or blemishes which differ from the substrate in their wettability. Assuming an amount of energy $W_{B}$ must be expended to pass a blemish, the overall energy required to displace a unit length contact line by $\Delta x$ is \cite{deGennes2004CapillarityPhenomena}:
\begin{equation} \label{blemish1}
F_{B} \Delta x = N W_{B}  \Delta x 
\end{equation}
where \textit{N} is the number of blemishes per unit area of the flat solid (i.e. number density). Consequently, on a surface with the solid surface fraction of $\phi_{s}$, Equation (\ref{blemish1}) is modified to:
\begin{equation} \label{blemish2}
F_{B} \Delta x = \phi_{s} N W_{B}  \Delta x 
\end{equation}
For a given material, the pinning force due to the defects can be expressed as:
\begin{equation} \label{blemish3}
F_{B}  = b \phi_{s} 
\end{equation}
where \textit{b} is a constant value for the flat and structured surfaces of the same surface chemistry. It should be noted that there probably exist more than one kind of blemish on a real surface, which are all included in Equation (\ref{blemish3}).

\subsection{Adaptation, $F_{A}$}
Although the defect model can explain pinning on some surfaces, the contact lines get pinned (i.e. contact angle hysteresis does exist) even on smooth and defect-free surfaces such as liquid-like coatings \cite{Chen1991MolecularSurfaces,Wang2016CovalentlyRepellency,Chen2023OmniphobicSurfaces}. This phenomenon has been attributed to the surface molecular re-orientation or the adaptation of the solid surface to the liquid \cite{Chen1991MolecularSurfaces, Butt2018AdaptiveWetting, Tadmor2021OpenForces}. Butt and co-workers have proposed a model to describe how the contact between the liquid and substrate changes the interfacial tensions and local surface wettability \cite{Butt2018AdaptiveWetting}. Although re-orientation or adaptation happens everywhere that the liquid is in contact with the solid, its intensity is much larger in the close vicinity of the three-phase contact line due to the pressure singularity at the contact line arising from differences in Laplace pressure \cite{Butt2018AdaptiveWetting, Tadmor2021OpenForces, Wong2020AdaptivePolydimethylsiloxane}. Here we assume that the effect of adaptation is dominant in a narrow region right under the microscopic three-phase contact line (i.e. the $\delta$-region), and the blemish model is dominant outside that region. The width of the $\delta$-region was estimated to be between 10 and 100 nm \cite{Butt2018AdaptiveWetting, Li2021AdaptationWater}, which is negligible compared to the size of the microstructures.  According to the general Equation (\ref{general}) for contact lines of unit length, one could write:
\begin{equation} \label{adap1}
F_{A} \Delta x  = \left(\frac{L_{TCL}}{L}\right)W_{A}  \Delta x = \delta W_{A}  \Delta x 
\end{equation}
where $W_{A}$ is the energy required for the contact line to pass a unit area of the $\delta$-region. Again, assuming that $W_{A}$ and the width of the $\delta$-region are topography-independent, the pinning force due to molecular re-orientation/adaptation simplifies to:
\renewcommand{\thefootnote}{\roman{footnote}}
\begin{equation} \label{adap2}
F_{A} = a \delta 
\end{equation}

\noindent where \textit{a} is assumed to be a constant value for the flat and structured surfaces of the same surface chemistry. The normal component of the surface tension force acting on the solid is balanced by surface de-formation in the case of a flexible surface and/or molecular re-orientation in the case of a perfectly rigid surface \cite{Chen1991MolecularSurfaces}. The value of $a$, or the intensity of the molecular re-orientation or adaptation, is, therefore, proportional to the normal component of the force applied on the solid material, i.e. $\gamma \sin \theta_{m}$, where $\theta_{m}$ is the microscopic contact angle between the liquid and solid material \cite{Chen1991MolecularSurfaces,Tadmor2009MeasurementSubstrate, Tadmor2021OpenForces}. However, considering that the microscopic contact angle on the solid is somewhere between the intrinsic advancing and receding contact angles, the value of $\theta_{m}$ has been assumed to be equal to the equilibrium contact angle on the solid material, and therefore, depends solely on the surface chemistry. It should be noted that adaptation, as well as the blemishes, can cause different pinning forces for the advancing and receding contact lines. While such differences affect the actual values of \textit{a} and \textit{b}, the functional forms of the Equations (\ref{blemish3}) and (\ref{adap2}) remain unaffected. These functions are summations of all types of pinning sources (as in Equation (\ref{general})), and as general models, the aforementioned differences are incorporated in them. Although in Equation (\ref{adap2}) the specific geometry of the microstructures is not taken into account, it should be noted that it could have an effect on molecular re-orientation by changing the speed of the microscopic contact line displacement \cite{Chen1991MolecularSurfaces,Butt2018AdaptiveWetting}. We will discuss this further below.

According to Butt and co-workers, the \emph{dynamic} advancing and receding contact angles on an adaptive interface are velocity-dependent as long as the motion is not too fast compared to the adaptation process \cite{Butt2018AdaptiveWetting}. Therefore, the dynamic advancing and dynamic receding contact angles can be expressed as functions of the contact line velocity \cite{Butt2018AdaptiveWetting}:

 \begin{eqnarray} \label{CA_VF}
\cos \theta_{a} (v_{a}) = \lim_{v_{a}\to0} \cos\theta_{a} - \frac{\Delta \gamma_{SL}}{\gamma} g(v_{a}) 
\end{eqnarray}
and
\begin{eqnarray} \label{CA_VB}
\cos \theta_{r} (v_{r}) = \lim_{v_{r}\to0} \cos\theta_{r} + \frac{\Delta \gamma_{S}}{\gamma} h(v_{r}) 
\end{eqnarray}

\noindent where $\Delta\gamma_{SL}$ and $\Delta \gamma_{S}$ are the maximum change in solid-liquid interfacial tension and the maximum change in solid-air interfacial tension due to adaptation, respectively. In the above equations, $v_{a}$ and $v_{r}$ stand for the velocities of the advancing and receding contact lines, and $\lim_{v_{a}\to\infty} g(v_{a}) = 1$ and $\lim_{v_{r}\to\infty} h(v_{r}) = 1$ where both describe first-order relaxation processes. In the model by Butt and co-workers, the advancing contact line is affected by the adaptation of a dry part of the solid to the liquid, whereas the recovery of a pre-wetted area affects the receding contact line \cite{Butt2018AdaptiveWetting}. This is the reason for different interfacial tensions (i.e. $\Delta\gamma_{SL}$ and $\Delta\gamma_{S}$) to appear in Equations (\ref{CA_VF}) and (\ref{CA_VB}). In this paper, we will be focusing on the range of the force required to move the contact lines on an adaptive surface. According to Equations (\ref{CA_VF}) and (\ref{CA_VB}), the dynamic contact angles will go to a constant value at high velocities and become velocity-independent, indicating the surface is not given enough time to adapt. Hence, $F^{max}_{d} = \gamma(\lim_{v_{r}\to\infty}\cos\theta_{r}-\lim_{v_{a}\to\infty}\cos\theta_{a})$ is the force required to move the unit-long contact lines from a perfectly adapted region to a completely unadapted one.

\subsection{Total de-pinning force, $F_{d}$}
According to Equation (\ref{general}), the total de-pinning force is the summation of the de-pinning forces due to different pinning mechanisms: defects (blemishes) and molecular re-orientation. Hence, combining Equations (\ref{general}), (\ref{blemish3}) and (\ref{adap2}), the total de-pinning force on a surface can be described as:
\begin{eqnarray} \label{Fdtotal}
F_{d}= F_{B} + F_{A} = b\phi_{s} + a\delta
\end{eqnarray}

In this general model, $b$ and $a$ have the same units as surface tension. They are equal to or larger than 0, and indicate the relative importance of each pinning mechanism. Based on Equation (\ref{Fdtotal}), a small contact angle hysteresis on a flat surface is likely to be an indication of a small value of $b$. The reason is that in this case, the largest possible area of the solid-liquid contact (i.e. $\phi_{s}=1$) has resulted in a small pinning force, which could be reflecting a nearly defect-free surface. The relationship between $b$ and $a$ can be complicated; for instance, some blemishes can be more adaptive than others. Furthermore, we note that Equation (\ref{eq_delta}) may not be an accurate expression for significantly anisotropic arrays. Hence, the relation to calculate the density of the $\delta$-region can be modified to make the model accurate for such special cases; for example, when the surface is decorated with microstripes. 

It should be noted that normally in studies focusing on microstructured surfaces, each of the $\delta$ and $\phi_{s}$ of the superhydrophobic designs can be defined as a monotonic function of another (see Figure \ref{fig_litData}a). Consequently, as can be seen in Figure \ref{fig_litData}, by excluding the flat surface, the de-pinning force data on each set of microstructured surfaces are correlated with both $\delta$ and $\phi_{s}$. To address this limitation, we take advantage of the reported data by Priest et al. \cite{Priest2009AsymmetricSurfaces} consisting of microholes and microposts, as well as using a defect-free liquid like coating on our designed surfaces to test our model Equation (\ref{Fdtotal}). It is worth mentioning that the defect (or blemish) model cannot practically be quantified on the actual surfaces \cite{Tadmor2021OpenForces} as the parameters of Equation (\ref{blemish1}) are too challenging to obtain experimentally, whereas the adaptation model \cite{Butt2018AdaptiveWetting} provides a measurable description of the contact line pinning (see Equations (\ref{CA_VF}) and (\ref{CA_VB})). Therefore, by using a defect-free liquid-like coating and suppressing the contribution of blemishes, the pinning mechanism that remains, i.e. adaptation,  is quantifiable to serve the purpose of model validation. 

In the following sections we focus experimentally on the case of flat and micropillared surfaces covered with a low-hysteresis QLS coating and discuss how the adaptation theory can explain pinning. We then fit the model Equation (\ref{Fdtotal}) to our experimental data and the data from literature and show that following this approach, contact line pinning and hysteresis on flat and microstructured surfaces can be explained within one theoretical framework.

\section{Experimental section}

\subsection{Surface fabrication}
Microstructured surface fabrication was done within the Class 10 cleanrooms of the Scottish Microelectronic Centre (SMC) using an etching process to pattern the substrate \cite{Zhao2020DropletSurfaces, AlBalushi2022BinarySurfaces}. The full description of the designed geometries is provided in Supplementary Information. The 4-inch silicon wafer (Silicon Materials, Landsberg, Germany) was thermally oxidised using a wet oxidation process at $1100^{\;\circ}$C to achieve a 0.3 $\mu$m layer of oxide. The wafer was then plasma-cleaned for 30 minutes in $\rm{O_{2}}$. Immediately after the plasma cleaning, the wafer was treated for 10 minutes at room temperature in Hexamethyldisilazane (HMDS, Thermo Fisher Scientific) to ensure the adhesion between the silicon surface and photoresist. The diluted negative photoresist nLoF2070 (MicroChemicals GmbH) was spin-coated onto the wafer using the manual POLOS spinner. The process used has two steps: 1) spread the photoresist for 10 s at 700 rpm, 2) spin the photoresist for 1 min at 3000 rpm. The photoresist was soft-baked on the hot plate for 1.5 min at $90^{\;\circ}$C. The thickness of the photoresist was measured using the reflectometer Nanospec: 0.65 $\mu$m. The photoresist-coated wafer was exposed to 365 nm wavelength UV light provided by LED (120 $\rm{mJ.cm^{-2}}$) using the mask-less photolithography machine (MicroWrite ML3 Pro, Durham Magneto Optics Ltd.), followed by the post-exposure bake for 1 min at $110^{\;\circ}$C. It was then developed using AZ 726 Developer (Merck) for 2 min with mild agitation. Then it was rinsed with de-ionised water and dried using a nitrogen gun. The wafer was inspected, using a Leica optical microscope, to make sure that all the patterns were resolved in photoresist. The oxide was dry ethched using JLS RIE, using $\rm{CHF_{3}}$ and $\rm{O_{2}}$ chemistry. The silicon was etched (with a Bosch process) in the DRIE system STS Multiplex ICP (inductively coupled plasma) for 105 cycles (12 seconds etch, 8 seconds passivation). After etching, the mask was removed by immersing the wafer in MICROPOSIT Remover 1165 (Dupont) at $60^{\;\circ}$C. The wafer was then rinsed with 2-propanol (Thermo Fisher Scientific) and then deionised water, before etching off the oxide layer. The oxide was etched using buffered hydrofluoric etch (BHF) 4:1 ($\rm{NH_{4}F}$, HF) (A-Gas Electronic Materials). The wafer was rinsed with de-ionised water and dried using a nitrogen gun. Finally, the wafer was inspected optically, using the Leica microscope and with the SEM (Tescan Mira).

\subsection{Coating optimisation}
 We used a QLS coating, consisting of Polydimethylsiloxane (PDMS) chains covalently grafted onto the substrate (i.e. silicon wafer) to coat the surfaces \cite{Khatir2020DesignSubstrate}. The chemical reaction was done in the vapour phase to achieve a uniform and conformal layer on the surfaces. To coat the silicon wafer with QLS, the substrate was first cleaned by rinsing with toluene (Sigma Aldrich) and 2-propanol (Sigma Aldrich). The substrate was then air-dried and placed in the oxygen plasma oven (HPT-200, Henniker Scientific Ltd.) to create active sites on the surface \cite{Khatir2020DesignSubstrate}. After plasma treatment, the substrate and 1,3-dichlorotetramethyldisiloxane (Thermo Fisher Scientific) were placed in a glass dish without direct contact (i.e. the substrate was fixed inside the lid) \cite{Khatir2020DesignSubstrate}. We used 3 $\mu$L of 1,3-dichlorotetramethyldisiloxane per $\rm{cm^{3}}$ of the dish volume. The dish was kept closed for 5 minutes and then the substrate was removed, rinsed with 2-propanol and toluene, and air-dried \cite{Khatir2020DesignSubstrate}. The reaction was done in ambient temperature (i.e. $20-25^{\;\circ}$C) and humidity (i.e. $30-40\%$ RH). Since contact angle hysteresis on liquid-like coatings is a function of the reduced grafting density \cite{Gresham2023NanostructureSurfaces}, changing the time and power of the plasma treatment while keeping the reaction time constant, is a strategy to adjust hysteresis. Therefore, to achieve an extremely low hysteresis, we studied the contact angle hysteresis as a function of plasma time and power, and tried different plasma settings to achieve a contact angle hysteresis of around $1^{\circ}$. Figure S3 in Supplementary Information shows how the advancing and receding contact angles on a flat silicon wafer changed with different plasma settings. As provided in Figure S3, 25 minutes of plasma treatment at 60 W power resulted in the lowest average hysteresis on a flat silicon wafer. Hence, we set the plasma time and power equal to these values to coat the flat and micropillared samples. The static advancing and static receding contact angles on the flat, unetched part of the same wafer were measured as $\theta^{static}_{a}=103.0 \pm 0.1 ^{\circ}$ and $\theta^{static}_{r}=101.7 \pm 0.3 ^{\circ}$ using the volume-change method \cite{Huhtamaki2018Surface-wettingMeasurements}.

\subsection{Surface characterisation}
The volume-change measurements were used only for the coating optimisation, and taken using a drop shape analyser setup. The sample was held horizontally on a triple-axis stage (PT3/M, Thorlabs). A 4-$\mu$L water droplet was placed on the surface using a programmable microfluidic syringe pump (ExiGo, Cellix) and left to relax for 10 s. The droplet was then inflated or deflated at $\pm 0.05 \; \mu \rm{L.s^{-1}}$. The droplet side profile was recorded using a camera (uEye 2D, IDS Imaging) during the experiment. The videos of the volume-change experiments (used for coating optimisation) were analysed using an open-source package, PyDSA \cite{Launay2018PyDSA:Python}. The static advancing ($\theta^{static}_{a}$) and static receding ($\theta^{static}_{r}$) contact angles are the largest and smallest contact angles before the contact line advances or recedes, respectively.

The tilted-plane experiments were carried out using a tilting-stage drop shape analyser (DSA100, KRÜSS Scientific) equipped with a light source and camera. The studied surfaces were fixed perpendicularly to the camera viewing direction. Hence, the front and back contact lines are expected to be in focus simultaneously in a plane perpendicular to the camera viewing direction.  In all the experiments following this method, water droplets of $8.5 \pm 0.6 \; \mu$L volume were used. Based on our preliminary observations, we estimated the tilt angle at which the droplets started and kept moving, $\omega_{s}$. To collect the data given in Figure \ref{fig_pillarDesign}d, the stage was initially tilted by an angle $\omega_{1} < \omega_{s}$ to reduce the total experiment time. This way, all the experiments in Figure \ref{fig_pillarDesign}d  took roughly the same time. After that initial tilting, the stage was further tilted by $1^{\circ}$ at each step and the droplet was allowed to relax before the next step. We set the rest time to 12 s, which is longer than the contact line relaxation time on a low-hysteresis liquid-like coating made of PDMS \cite{Barrio-Zhang2020Contact-AngleSurfaces}. To collect the data in Figure \ref{fig_CA_V}, the stage was tilted at 0.4 $\rm{deg.s^{-1}}$ until the droplet moved out of the camera frame. The full stage tilting curves used to collect the data are provided in Figure S4a of the Supplementary Information.

We developed a piece of code in MATLAB and used it to analyse the videos of the tilting plane experiments. In both volume-change and tilting methods, Canny algorithm \cite{Canny1986ADetection} was used for edge detection, and as has previously been suggested \cite{Andersen2017DropMethod}, the droplet boundary was divided into left and right parts, and two elliptical curves were fitted \cite{ChernovEllipseMethod} to find the contact angles. Since droplets on a superhydrophobic surface can sometimes slightly roll out of the plane, the baseline was detected in each frame. To do this, the points on the lower 1/5 of the droplet left and right-hand sides with the smallest distance in between were found. Since we were working with contact angles larger than 90$^{\circ}$, those points were considered to be located on the baseline. Figure S4b in Supplementary Information shows these steps. 

\section{Results and discussion}

\begin{figure}[h!]
  \begin{center}
  
  \vspace{-0.28 cm}
    \includegraphics[width=1\textwidth]{ 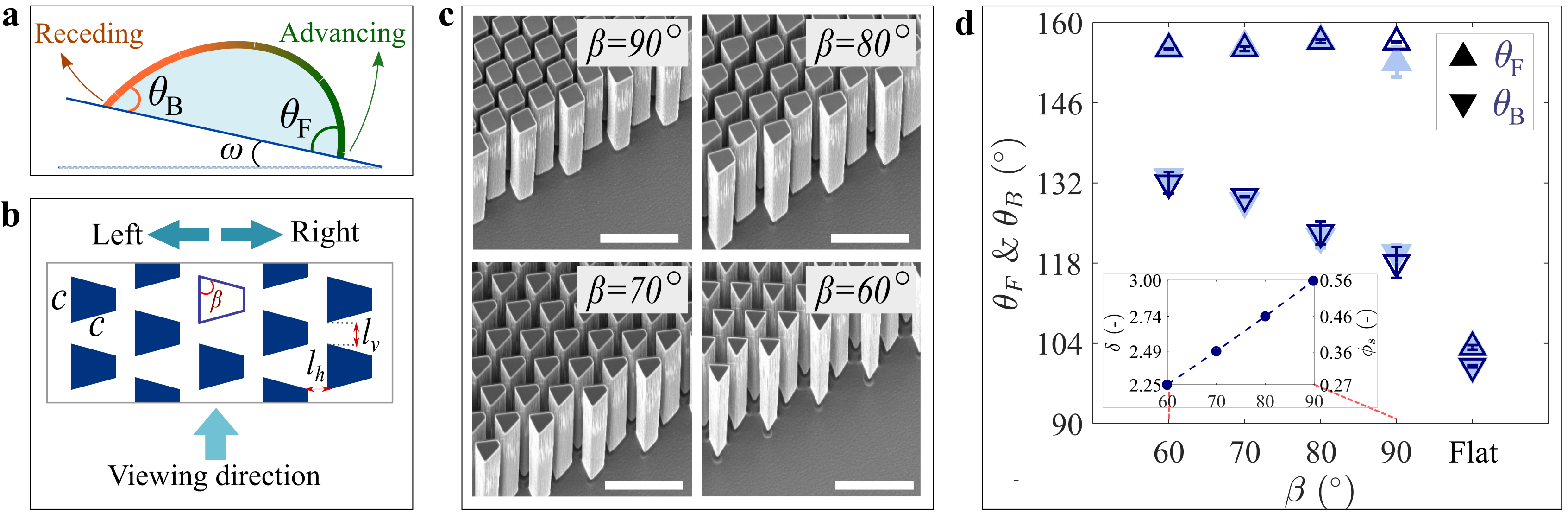}
    \vspace{-0.65 cm}
  \end{center}
\caption{Surface design and results based on the tilting method: \textbf{a)} a droplet on a surface at the tilt angle of $\omega$ with the front and back contact angles $\theta_{F}$ and $\theta_{B}$. \textbf{b)} \textit{c} and $\beta$ are the parameters that define the pillar geometry. The triangular array is described by the vertical gap ($l_{v}$) and the horizontal gap ($l_{h}$) between the pillars. Left and Right indicate the direction of droplet motion with respect to the pillars. \textbf{c)} SEM images of the silicon pillars; scale bars are 50 $\mu$m. \textbf{d)} Measured values of $\theta_{F}$ and $\theta_{B}$ on flat and micropillared surfaces along with the corresponding $\phi_{s}$ and $\delta$ in the inset. Open and filled triangles show the direction of droplet motion to the left and right, respectively. Error bars present the sample standard deviation for three independent experiments.}
  \label{fig_pillarDesign}
\end{figure}

To calculate the pinning force on superhydrophobic surfaces, four different pillar geometries, from square to triangle, were arranged in triangular arrays, as detailed in the Supplementary Information. The design of the surface can be fully described using 5 parameters. A top-view of an array design is illustrated in Figure \ref{fig_pillarDesign}b. We set $c = 15 \; \mu$m, and $l_{v} = l_{h} = 5 \; \mu$m in all the designs. The pillar height $H \approx 50 \; \mu$m and the only varying parameter was $\beta$.  
Contact angle hysteresis on liquid-like coatings changes with the reduced grafting density \cite{Gresham2023NanostructureSurfaces}, which is an indication of the chain stretching \cite{Brittain2007ABrushes}. The reduced grafting density is equal to $\sigma \pi R_{g}^{2}$ where $\sigma$ and $R_{g}$ are the grafting density and the radius of gyration of the grafted polymer, respectively \cite{Brittain2007ABrushes}. Low-hysteresis liquid-like coatings made of PDMS have been reported to have a reduced grafting density between 1 and 3 \cite{Gresham2023NanostructureSurfaces, Gresham2023AdvancesSurfaces}, suggesting the grafted layer is in the transition between mushroom and brush regimes \cite{Brittain2007ABrushes}. This means the grafted chains have enough space to bend and rotate, while packed well enough to cover the substrate and smooth out the physical or chemical defects \cite{Chen2023OmniphobicSurfaces} — both of these characteristics result in low contact angle hysteresis. For the combination of this surface chemistry (with water contact angle larger than $100^{\circ}$) and our designed geometries, the droplets are predicted to be in the suspended state \cite{Lafuma2003SuperhydrophobicStates} (see Supplementary Information for calculations), which is consistent with our experimental observations. Since the pillar geometry is not symmetric in all the arrays, the droplet de-pinning was studied in two directions, left and right, as defined in Figure \ref{fig_pillarDesign}b.

We used the tilting plane method to measure the smallest difference between front (i.e. the advancing side) and back (i.e. the receding side) contact angles that makes a droplet start and keep sliding. The method is detailed in the previous section and Supplementary Information. We refer to our measured advancing contact angles as $\theta_{F}$ and to the measured receding contact angles as $\theta_{B}$ to distinguish between the tilting method as a dynamic method and the static methods of measurements \cite{Krasovitski2005DropsPlate}. The front and back contact angles at which the $8.5 \pm 0.6 \;\mu$L water droplets began and kept sliding are given in Figure \ref{fig_pillarDesign}d.  

\begin{figure}[b!]
\begin{center}
\includegraphics[width=0.8\linewidth]{ 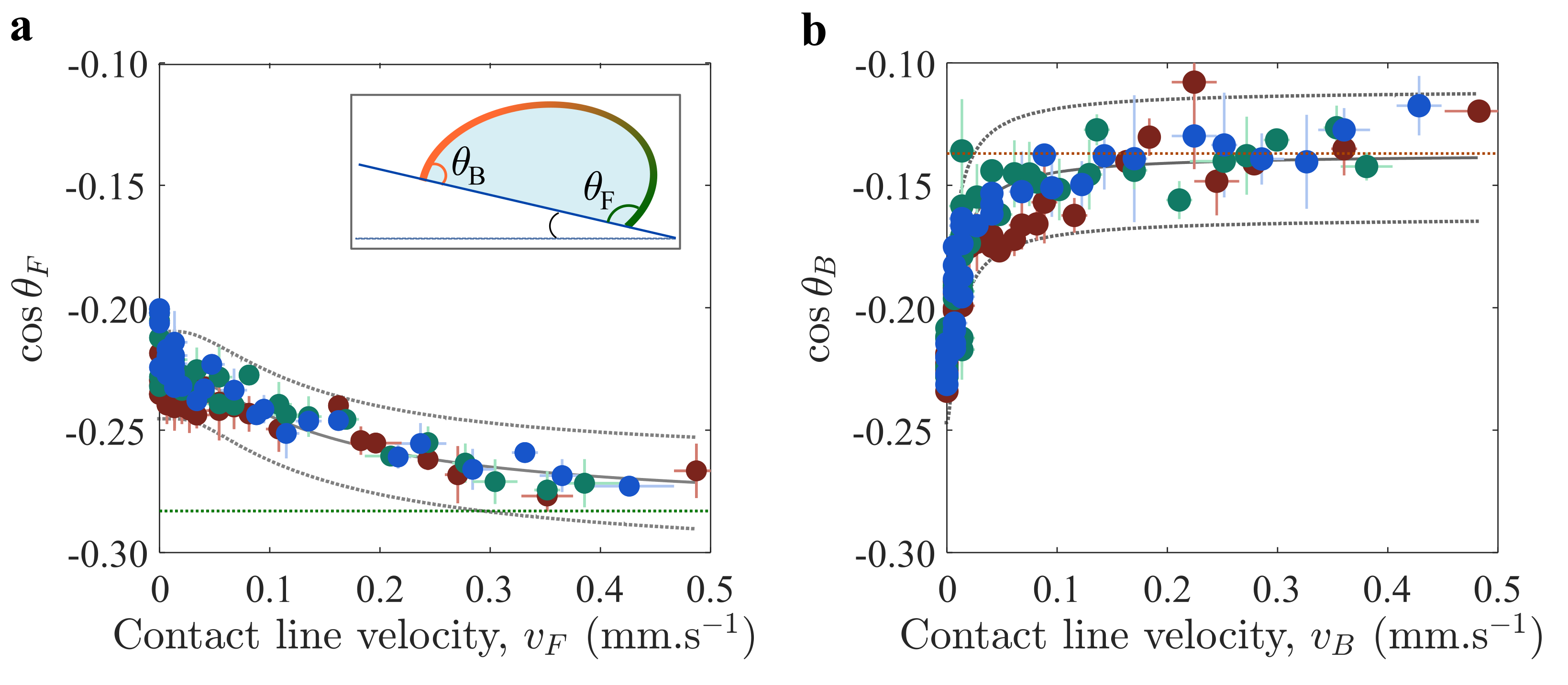}
\caption{The dependence of the dynamic contact angles on contact line velocity on the flat reference surface: \textbf{a)} the front (advancing) contact line; \textbf{b)} the back (receding) contact line. Different colours present independent experiments, and the stage was tilted at 0.4 $\rm{deg.s^{-1}}$ in all the experiments. Data show averages of 10 data points and error bars present their standard deviations. Gray curves show fitted Equations (\ref{CA_VF}) and (\ref{CA_VB}) along with their 95\% confidence bounds. Dotted horizontal lines present the predicted limiting plateau values based on the fitted equations.  }
    \label{fig_CA_V}
\end{center}
    
\end{figure}

The experimental data given in Figure \ref{fig_pillarDesign}d  suggest that the force required to move the droplet (or the de-pinning force according to Equation (\ref{Fd})) increases from $\beta=60^{\circ}$ to $\beta=90^{\circ}$. The results also show that the direction of droplet motion has not made a statistically significant difference in the contact angles. 
The effect of $\phi_{s}$ on $F_{d}$ calculated using the data in Figure \ref{fig_pillarDesign}d was not statistically significant (p-value $\approx$ 0.4). This indicates that the blemishes do not determine the pinning force on these surfaces, which is consistent with the properties of a low-hysteresis liquid-like coating \cite{Chen2023OmniphobicSurfaces}. For water with $\gamma=72 \;\rm{mN.m^{-1}}$, fitting Equation (\ref{Fdtotal}) to the data corresponding to Figure \ref{fig_pillarDesign}d results in $a=9.02 \pm 1.08\;\rm{mN.m^{-1}}$, with \textit{b} set to zero. We now compare the fitted value of \textit{a} to the range of predictions based on the adaptation model \cite{Butt2018AdaptiveWetting}. Since both flat and structured surfaces have the same surface chemistry, the fitted value of \textit{a} is expected to be compatible with the range of predictions based on Equations (\ref{CA_VF}) and (\ref{CA_VB}).

 To find the limiting contact angles according to Equations (\ref{CA_VF}) and (\ref{CA_VB}), a water droplet was placed on the flat QLS-coated reference surface, and the stage was tilted at 0.4 $\rm{deg.s^{-1}}$ until the droplet moved out of the camera frame. Figure \ref{fig_CA_V} shows the change of contact angle with velocity , where each reported value presents the average of 10 data points, along with the 95\% confidence bounds based on Equations (\ref{CA_VF}) and (\ref{CA_VB}). Based on these data, the limiting values of $\cos \theta_{F}$ and $\cos \theta_{B}$ at large contact line velocities (i.e. velocity $\rightarrow \; \infty$) are equal to $-0.283 \pm 0.013$ and $-0.137 \pm 0.014$, respectively. Hence, considering the contact lines facing the same length of the $\delta$-region on a flat surface (i.e. $\delta=1$), based on the limiting contact angles, the largest pinning force for water is $F^{max}_{d} = \gamma(\cos\theta^{min}_{B}-\cos\theta_{F}^{max}) = a^{max} = 10.51 \pm 1.38\; \rm{mN.m^{-1}}$.\\
 Although the pinning force due to adaptation depends on the microscopic velocity of the contact line, the value of the $F^{max}_{d}$ provides an estimation of the strength of the $\delta$-region on our surfaces. It can be seen that this value is compatible with the fitted value of $a=9.02 \pm 1.08 \; \rm{ mN.m^{-1}}$, suggesting adaptation as a reasonable explanation for pinning on these surfaces. 
 
\begin{figure}[h]
  \begin{center}
  
  \vspace{-0.18 cm}
    \includegraphics[width=0.85\textwidth]{ 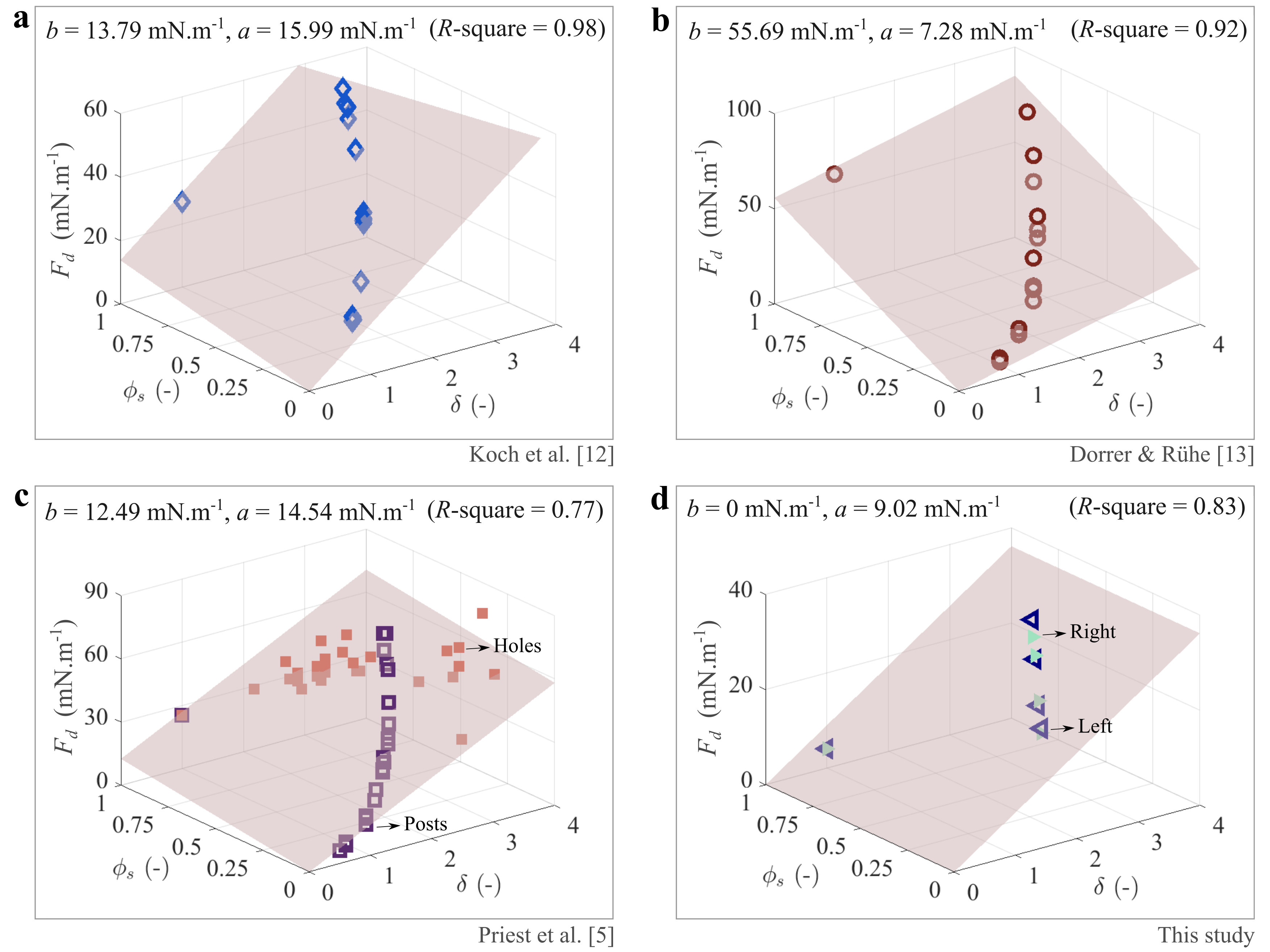}
    \vspace{-0.75 cm}
  \end{center}
\caption{Comparison of the model against experimental data: the pink plane shows Equation (\ref{Fdtotal}) fitted to the de-pinning force data by \textbf{a)} Koch et al. \cite{Koch2014ModelingSurfaces}; \textbf{b)} Dorrer and Rühe \cite{Dorrer2006AdvancingSurfaces}; \textbf{c)} Priest et al. \cite{Priest2009AsymmetricSurfaces}; and \textbf{d)} our experiments. All the values of $F_{d}$ are calculated according to the Equation (\ref{Fd}) and based on the experimental data for the advancing and receding contact angles.}
  \label{fig_litFitted}
\end{figure}

To further test our model, we fitted the model Equation (\ref{Fdtotal}) to our experimental results and the data from literature that were earlier shown in Figure \ref{fig_litData}. The fitted planes and values of \textit{a} and \textit{b} can be seen in Figure \ref{fig_litFitted}. Some imperfections, such as added roughness on the edges, are inevitable in the fabrication process, and are most likely different in the arrays of posts and arrays of holes — reducing the accuracy of the predictions. Nonetheless, as Figure \ref{fig_litFitted} shows, for a given surface chemistry, the de-pinning force can still be described as a function of $\phi_{s}$ and $\delta$, which are determined by the design of the microstructures.

In fitting Equation (\ref{Fdtotal}) to the data reported by Priest et al. \cite{Priest2009AsymmetricSurfaces}, Koch et al. \cite{Koch2014ModelingSurfaces} and Dorrer and Rühe \cite{Dorrer2006AdvancingSurfaces}, the effect of both $\phi_{s}$ and $\delta$ was statistically significant. 
In all these studies, the chemicals used to hydrophobise the surfaces were polymeric molecules with different functional groups \cite{Priest2009AsymmetricSurfaces, Koch2014ModelingSurfaces, Dorrer2006AdvancingSurfaces}. Such surface molecules have internal degrees of freedom that allows them to re-orient so that those groups that are more hydrophilic would be in contact with water, which explains why molecular re-orientation plays a role in all of them (i.e. $a > 0$) \cite{Tadmor2021OpenForces}. It can be seen in Figure \ref{fig_litFitted} that the value of \textit{b} fitted to the data by Dorrer and Rühe is much larger than \textit{a}. This indicates the stronger effect of surface heterogeneity in giving rise to contact line pinning. According to their report, they used a surface treatment process in the liquid phase \cite{Dorrer2006AdvancingSurfaces}. Liquid film processes, however, could sometimes result in non-uniform layers due to the effects such as dry spot or pattern formation during drying \cite{Orell1971FormationBelow, Schwartz2001DewettingFilm}. They have mentioned the coating thickness to be between 10 and 15 nm \cite{Dorrer2006AdvancingSurfaces}, potentially implying the relative non-uniformity of the surface. The large fitted value of \textit{b} could thus be a reflection of the process they used. In the other two studies, by Koch et al. \cite{Koch2014ModelingSurfaces} and Priest et al. \cite{Priest2009AsymmetricSurfaces}, the fitted values of \textit{a} and \textit{b} are comparable. Although Priest et al. also used a treatment process in the liquid phase, the process that they have used was reported to form an even coating as much as could be revealed through SEM imaging \cite{Shirtcliffe2004ThePrototyping}. This could explain why the effect of defects or blemishes has remained relatively small on the surfaces prepared by Priest et al. \cite{Priest2009AsymmetricSurfaces}.

\section{Conclusions}
Neither the solid surface fraction ($\phi_{s}$) nor the normalized length of the maximal three-phase contact line ($\delta$) on their own can unify the (de-)pinning force on both flat and superhydrophobic surfaces. We suggest that the de-pinning (or hysteresis) force is a function of both $\phi_{s}$ and $\delta$ on flat surfaces, arrays of posts and arrays of holes. The model equation was compared against experimental data. We attribute the contribution of $\phi_{s}$ to the blemishes (or defects) on the solid material \cite{Joanny1984AHysteresis}, and the contribution of $\delta$ to molecular re-orientation or adaptation \cite{Chen1991MolecularSurfaces, Butt2018AdaptiveWetting} as different mechanisms for contact line pinning that can have a complex relationship with each other and with the de-pinning speed. This theoretical framework provides insight into the interplay between the topography, surface chemistry and different pinning mechanisms, and can be used in designing micro-engineered surfaces on which the behaviour of the contact line is important. This approach also highlights surface structuring as a potential strategy to quantitatively distinguish the role of blemishes or defects from that of molecular re-orientation in giving rise to contact line pinning on a given solid material.
\\
\newline

\noindent\textbf{\Large{CRediT authorship contribution statement}} \\

\noindent \textbf{M. Meyari:} Conceptualization, Methodology, Software, Validation, Formal Analysis, Investigation, Visualization, Writing (original draft), Writing (review \& editing); \textbf{C. Dunare:} Visualization, Resources, Writing (review \& editing); \textbf{K. Sefiane:} Methodology, Supervision, Funding Acquisition, Writing (review \& editing); \textbf{S. Titmuss:} Methodology, Supervision, Writing (review \& editing); \textbf{J. Thijssen:} Methodology, Supervision, Project Administration, Writing (review \& editing)
\newline
\\

\noindent\textbf{\Large{Acknowledgements}} \\

\noindent M.M. wishes to thank Hernán Barrio-Zhang for valuable discussions. M.M. also acknowledges funding from the EPSRC Centre for Doctoral Training in Soft Matter for Formulation and Industrial Innovation [SOFI2-CDT, EP/S023631/1]. 
For the purpose of open access, the authors have applied a Creative Commons Attribution (CC BY) licence to any Author Accepted Manuscript version arising from this submission.
\newline
\\
\\
\noindent\textbf{\Large{Data availability}} \\
\noindent The data that support the findings of this study are available on Edinburgh DataShare.


\begin{thebibliography}{10}
\providecommand{\url}[1]{#1}
\csname url@samestyle\endcsname
\providecommand{\newblock}{\relax}
\providecommand{\bibinfo}[2]{#2}
\providecommand{\BIBentrySTDinterwordspacing}{\spaceskip=0pt\relax}
\providecommand{\BIBentryALTinterwordstretchfactor}{4}
\providecommand{\BIBentryALTinterwordspacing}{\spaceskip=\fontdimen2\font plus
\BIBentryALTinterwordstretchfactor\fontdimen3\font minus \fontdimen4\font\relax}
\providecommand{\BIBforeignlanguage}[2]{{%
\expandafter\ifx\csname l@#1\endcsname\relax
\typeout{** WARNING: IEEEtran.bst: No hyphenation pattern has been}%
\typeout{** loaded for the language `#1'. Using the pattern for}%
\typeout{** the default language instead.}%
\else
\language=\csname l@#1\endcsname
\fi
#2}}
\providecommand{\BIBdecl}{\relax}
\BIBdecl

\bibitem{Bonn2009WettingSpreading}
\BIBentryALTinterwordspacing
D.~Bonn, J.~Eggers, J.~Indekeu, J.~Meunier, and E.~Rolley, ``{Wetting and spreading},'' \emph{Reviews of Modern Physics}, vol.~81, no.~2, pp. 739--805, 2009.   \url{https://dx.doi.org/10.1103/RevModPhys.81.739}
\BIBentrySTDinterwordspacing

\bibitem{Seo2014InfluenceDewing}
\BIBentryALTinterwordspacing
D.~Seo, C.~Lee, and Y.~Nam, ``{Influence of geometric patterns of microstructured superhydrophobic surfaces on water-harvesting performance via dewing},'' \emph{Langmuir}, vol.~30, no.~51, pp. 15\,468--15\,476, 2014.   \url{https://dx.doi.org/10.1021/LA5041486}
\BIBentrySTDinterwordspacing

\bibitem{Shanahan1995SimpleHysteresis}
\BIBentryALTinterwordspacing
M.~E.~R. Shanahan, ``{Simple Theory of "Stick-Slip" Wetting Hysteresis},'' \emph{Langmuir}, vol.~11, no.~2, pp. 1041--1043, 1995.   \url{https://dx.doi.org/10.1021/la00003a057}
\BIBentrySTDinterwordspacing

\bibitem{Butt2022ContactHysteresis}
\BIBentryALTinterwordspacing
H.~J. Butt, J.~Liu, K.~Koynov, B.~Straub, C.~Hinduja, I.~Roismann, R.~Berger, X.~Li, D.~Vollmer, W.~Steffen, and M.~Kappl, ``{Contact angle hysteresis},'' \emph{Current Opinion in Colloid {\&} Interface Science}, vol.~59, p. 101574, 2022.   \url{https://dx.doi.org/10.1016/J.COCIS.2022.101574}
\BIBentrySTDinterwordspacing

\bibitem{Long2006OnHysteresis}
\BIBentryALTinterwordspacing
J.~Long and P.~Chen, ``{On the role of energy barriers in determining contact angle hysteresis},'' \emph{Advances in Colloid and Interface Science}, vol. 127, no.~2, pp. 55--66, 2006.   \url{https://dx.doi.org/10.1016/J.CIS.2006.09.001}
\BIBentrySTDinterwordspacing

\bibitem{Cassie1944WettabilitySurfaces}
\BIBentryALTinterwordspacing
A.~B. Cassie and S.~Baxter, ``{Wettability of porous surfaces},'' \emph{Transactions of the Faraday Society}, vol.~40, pp. 546--551, 1944.   \url{https://dx.doi.org/10.1039/TF9444000546}
\BIBentrySTDinterwordspacing

\bibitem{Lafuma2003SuperhydrophobicStates}
\BIBentryALTinterwordspacing
A.~Lafuma and D.~Qu{\'{e}}r{\'{e}}, ``{Superhydrophobic states},'' \emph{Nature Materials}, vol.~2, no.~7, pp. 457--460, 2003.   \url{https://dx.doi.org/10.1038/nmat924}
\BIBentrySTDinterwordspacing

\bibitem{Priest2009AsymmetricSurfaces}
\BIBentryALTinterwordspacing
C.~Priest, T.~W.~J. Albrecht, R.~Sedev, and J.~Ralston, ``{Asymmetric wetting hysteresis on hydrophobic microstructured surfaces},'' \emph{Langmuir}, vol.~25, no.~10, pp. 5655--5660, 2009.   \url{https://dx.doi.org/10.1021/LA804246A}
\BIBentrySTDinterwordspacing

\bibitem{Cansoy2011EffectSurfaces}
\BIBentryALTinterwordspacing
C.~E. Cansoy, H.~Y. Erbil, O.~Akar, and T.~Akin, ``{Effect of pattern size and geometry on the use of Cassie–Baxter equation for superhydrophobic surfaces},'' \emph{Colloids and Surfaces A: Physicochemical and Engineering Aspects}, vol. 386, no. 1-3, pp. 116--124, 2011.   \url{https://dx.doi.org/10.1016/J.COLSURFA.2011.07.005}
\BIBentrySTDinterwordspacing

\bibitem{Qiao2017FrictionSurfaces}
\BIBentryALTinterwordspacing
S.~Qiao, S.~Li, Q.~Li, B.~Li, K.~Liu, and X.~Q. Feng, ``{Friction of Droplets Sliding on Microstructured Superhydrophobic Surfaces},'' \emph{Langmuir}, vol.~33, no.~47, pp. 13\,480--13\,489, 2017.   \url{https://dx.doi.org/10.1021/ACS.LANGMUIR.7B03087}
\BIBentrySTDinterwordspacing

\bibitem{Xu2012FromSurfaces}
\BIBentryALTinterwordspacing
W.~Xu and C.~H. Choi, ``{From sticky to slippery droplets: Dynamics of contact line depinning on superhydrophobic surfaces},'' \emph{Physical Review Letters}, vol. 109, no.~2, p. 024504, 2012.   \url{https://dx.doi.org/10.1103/PHYSREVLETT.109.024504}
\BIBentrySTDinterwordspacing

\bibitem{Extrand2002ModelSurfaces}
\BIBentryALTinterwordspacing
C.~W. Extrand, ``{Model for contact angles and hysteresis on rough and ultraphobic surfaces},'' \emph{Langmuir}, vol.~18, no.~21, pp. 7991--7999, 2002.   \url{https://dx.doi.org/10.1021/LA025769Z}
\BIBentrySTDinterwordspacing

\bibitem{Sarshar2019DepinningModels}
\BIBentryALTinterwordspacing
M.~A. Sarshar, Y.~Jiang, W.~Xu, and C.~H. Choi, ``{Depinning force of a receding droplet on pillared superhydrophobic surfaces: Analytical models},'' \emph{Journal of Colloid and Interface Science}, vol. 543, pp. 122--129, 2019.   \url{https://dx.doi.org/10.1016/J.JCIS.2019.02.042}
\BIBentrySTDinterwordspacing

\bibitem{Zhao2012Effectrobustness}
\BIBentryALTinterwordspacing
H.~Zhao, K.~C. Park, and K.~Y. Law, ``{Effect of surface texturing on superoleophobicity, contact angle hysteresis, and "robustness"},'' \emph{Langmuir}, vol.~28, no.~42, pp. 14\,925--14\,934, 2012.   \url{https://dx.doi.org/10.1021/LA302765T}
\BIBentrySTDinterwordspacing

\bibitem{Koch2014ModelingSurfaces}
\BIBentryALTinterwordspacing
B.~M. Koch, A.~Amirfazli, and J.~A. Elliott, ``{Modeling and measurement of contact angle hysteresis on textured high-contact-angle surfaces},'' \emph{Journal of Physical Chemistry C}, vol. 118, no.~32, pp. 18\,554--18\,563, 2014.   \url{https://dx.doi.org/10.1021/JP504891U}
\BIBentrySTDinterwordspacing

\bibitem{Dorrer2006AdvancingSurfaces}
\BIBentryALTinterwordspacing
C.~Dorrer and J.~Ruhe, ``{Advancing and receding motion of droplets on ultrahydrophobic post surfaces},'' \emph{Langmuir}, vol.~22, no.~18, pp. 7652--7657, 2006.   \url{https://dx.doi.org/10.1021/LA061452D}
\BIBentrySTDinterwordspacing

\bibitem{Jiang2019GeneralizedSurfaces}
\BIBentryALTinterwordspacing
Y.~Jiang, W.~Xu, M.~A. Sarshar, and C.~H. Choi, ``{Generalized models for advancing and receding contact angles of fakir droplets on pillared and pored surfaces},'' \emph{Journal of Colloid and Interface Science}, vol. 552, pp. 359--371, 2019.   \url{https://dx.doi.org/10.1016/J.JCIS.2019.05.053}
\BIBentrySTDinterwordspacing

\bibitem{deGennes2004CapillarityPhenomena}
\BIBentryALTinterwordspacing
P.-G. de~Gennes, F.~Brochard-Wyart, and D.~Qu{\'{e}}r{\'{e}}, \emph{{Capillarity and wetting phenomena}}.\hskip 1em plus 0.5em minus 0.4em\relax Springer New York, 2004.   \url{https://dx.doi.org/10.1007/978-0-387-21656-0}
\BIBentrySTDinterwordspacing

\bibitem{Joanny1984AHysteresis}
\BIBentryALTinterwordspacing
J.~F. Joanny and P.~G. De~Gennes, ``{A model for contact angle hysteresis},'' \emph{The Journal of Chemical Physics}, vol.~81, no.~1, pp. 552--562, 1984.   \url{https://dx.doi.org/10.1063/1.447337}
\BIBentrySTDinterwordspacing

\bibitem{Chen1991MolecularSurfaces}
\BIBentryALTinterwordspacing
Y.~L. Chen, C.~A. Helm, and J.~N. Israelachvili, ``{Molecular mechanisms associated with adhesion and contact angle hysteresis of monolayer surfaces},'' \emph{Journal of Physical Chemistry}, vol.~95, no.~26, pp. 10\,736--10\,747, 1991.   \url{https://dx.doi.org/10.1021/J100179A041}
\BIBentrySTDinterwordspacing

\bibitem{Butt2018AdaptiveWetting}
\BIBentryALTinterwordspacing
H.~J. Butt, R.~Berger, W.~Steffen, D.~Vollmer, and S.~A. Weber, ``{Adaptive wetting - adaptation in wetting},'' \emph{Langmuir}, vol.~34, no.~38, pp. 11\,292--11\,304, 2018.   \url{https://dx.doi.org/10.1021/ACS.LANGMUIR.8B01783}
\BIBentrySTDinterwordspacing

\bibitem{Tadmor2009MeasurementSubstrate}
\BIBentryALTinterwordspacing
R.~Tadmor, P.~Bahadur, A.~Leh, H.~E. N'Guessan, R.~Jaini, and L.~Dang, ``{Measurement of lateral adhesion forces at the interface between a liquid drop and a substrate},'' \emph{Physical Review Letters}, vol. 103, no.~26, p. 266101, 2009.   \url{https://dx.doi.org/10.1103/PHYSREVLETT.103.266101}
\BIBentrySTDinterwordspacing

\bibitem{Tadmor2021OpenForces}
\BIBentryALTinterwordspacing
R.~Tadmor, ``{Open Problems in Wetting Phenomena: Pinning Retention Forces},'' \emph{Langmuir}, vol.~37, no.~21, pp. 6357--6372, 2021.   \url{https://dx.doi.org/10.1021/ACS.LANGMUIR.0C02768}
\BIBentrySTDinterwordspacing

\bibitem{Wang2016CovalentlyRepellency}
\BIBentryALTinterwordspacing
L.~Wang and T.~J. Mccarthy, ``{Covalently Attached Liquids: Instant Omniphobic Surfaces with Unprecedented Repellency},'' \emph{Angewandte Chemie International Edition}, vol.~55, no.~1, pp. 244--248, 2016.   \url{https://dx.doi.org/10.1002/ANIE.201509385}
\BIBentrySTDinterwordspacing

\bibitem{Chen2023OmniphobicSurfaces}
\BIBentryALTinterwordspacing
L.~Chen, S.~Huang, R.~H. Ras, and X.~Tian, ``{Omniphobic liquid-like surfaces},'' \emph{Nature Reviews Chemistry}, vol.~7, no.~2, pp. 123--137, 2023.   \url{https://dx.doi.org/10.1038/s41570-022-00455-w}
\BIBentrySTDinterwordspacing

\bibitem{Wong2020AdaptivePolydimethylsiloxane}
\BIBentryALTinterwordspacing
W.~S.~Y. Wong, L.~Hauer, A.~Naga, A.~Kaltbeitzel, P.~Baumli, R.~Berger, M.~D'Acunzi, D.~Vollmer, and H.~J. Butt, ``{Adaptive Wetting of Polydimethylsiloxane},'' \emph{Langmuir}, vol.~36, no.~26, pp. 7236--7245, 2020.   \url{https://dx.doi.org/10.1021/ACS.LANGMUIR.0C00538}
\BIBentrySTDinterwordspacing

\bibitem{Li2021AdaptationWater}
\BIBentryALTinterwordspacing
X.~Li, S.~Silge, A.~Saal, G.~Kircher, K.~Koynov, R.~Berger, and H.~J. Butt, ``{Adaptation of a Styrene-Acrylic Acid Copolymer Surface to Water},'' \emph{Langmuir}, vol.~37, no.~4, pp. 1571--1577, 2021.   \url{https://dx.doi.org/10.1021/ACS.LANGMUIR.0C03226}
\BIBentrySTDinterwordspacing

\bibitem{Zhao2020DropletSurfaces}
\BIBentryALTinterwordspacing
H.~Zhao, D.~Orejon, C.~Mackenzie-Dover, P.~Valluri, M.~E. Shanahan, and K.~Sefiane, ``{Droplet motion on contrasting striated surfaces},'' \emph{Applied Physics Letters}, vol. 116, no.~25, 2020.   \url{https://dx.doi.org/10.1063/5.0009364}
\BIBentrySTDinterwordspacing

\bibitem{AlBalushi2022BinarySurfaces}
\BIBentryALTinterwordspacing
K.~M. Al~Balushi, K.~Sefiane, and D.~Orejon, ``{Binary mixture droplet wetting on micro-structure decorated surfaces},'' \emph{Journal of Colloid and Interface Science}, vol. 612, pp. 792--805, 2022.   \url{https://dx.doi.org/10.1016/J.JCIS.2021.12.171}
\BIBentrySTDinterwordspacing

\bibitem{Khatir2020DesignSubstrate}
\BIBentryALTinterwordspacing
B.~Khatir, S.~Shabanian, and K.~Golovin, ``{Design and High-Resolution Characterization of Silicon Wafer-like Omniphobic Liquid Layers Applicable to Any Substrate},'' \emph{ACS Applied Materials and Interfaces}, vol.~12, no.~28, pp. 31\,933--31\,939, 2020.   \url{https://dx.doi.org/10.1021/ACSAMI.0C06433}
\BIBentrySTDinterwordspacing

\bibitem{Gresham2023NanostructureSurfaces}
\BIBentryALTinterwordspacing
I.~J. Gresham, S.~G. Lilley, A.~R. Nelson, K.~Koynov, and C.~Neto, ``{Nanostructure Explains the Behavior of Slippery Covalently Attached Liquid Surfaces},'' \emph{Angewandte Chemie International Edition}, vol.~62, no.~41, p. e202308008, 2023.   \url{https://dx.doi.org/10.1002/ANIE.202308008}
\BIBentrySTDinterwordspacing

\bibitem{Huhtamaki2018Surface-wettingMeasurements}
\BIBentryALTinterwordspacing
T.~Huhtam{\"{a}}ki, X.~Tian, J.~T. Korhonen, and R.~H.~A. Ras, ``{Surface-wetting characterization using contact-angle measurements},'' \emph{Nature Protocols}, vol.~13, no.~7, pp. 1521--1538, 2018.   \url{https://dx.doi.org/10.1038/s41596-018-0003-z}
\BIBentrySTDinterwordspacing

\bibitem{Launay2018PyDSA:Python}
\BIBentryALTinterwordspacing
G.~Launay, ``{PyDSA: Drop Shape Analysis in Python},'' 2018.   \url{https://framagit.org/gabylaunay/pyDSA_core}  (accessed 02/13/2023)
\BIBentrySTDinterwordspacing

\bibitem{Barrio-Zhang2020Contact-AngleSurfaces}
\BIBentryALTinterwordspacing
H.~Barrio-Zhang, E.~Ruiz-Gutiérrez, S.~Armstrong, G.~McHale, G.~G. Wells, and R.~Ledesma-Aguilar, ``{Contact-Angle Hysteresis and Contact-Line Friction on Slippery Liquid-like Surfaces},'' \emph{Langmuir}, vol.~36, no.~49, pp. 15\,094--15\,101, 2020.   \url{https://dx.doi.org/10.1021/ACS.LANGMUIR.0C02668}
\BIBentrySTDinterwordspacing

\bibitem{Canny1986ADetection}
\BIBentryALTinterwordspacing
J.~Canny, ``{A Computational Approach to Edge Detection},'' \emph{IEEE Transactions on Pattern Analysis and Machine Intelligence}, vol. PAMI-8, no.~6, pp. 679--698, 1986.   \url{https://dx.doi.org/10.1109/TPAMI.1986.4767851}
\BIBentrySTDinterwordspacing

\bibitem{Andersen2017DropMethod}
\BIBentryALTinterwordspacing
N.~K. Andersen and R.~Taboryski, ``{Drop shape analysis for determination of dynamic contact angles by double sided elliptical fitting method},'' \emph{Measurement Science and Technology}, vol.~28, no.~4, p. 047003, 2017.   \url{https://dx.doi.org/10.1088/1361-6501/AA5DCF}
\BIBentrySTDinterwordspacing

\bibitem{ChernovEllipseMethod}
\BIBentryALTinterwordspacing
N.~Chernov, ``{Ellipse Fit (Taubin method)}.''   \url{https://uk.mathworks.com/matlabcentral/fileexchange/22683-ellipse-fit-taubin-method} (accessed 05/09/2024)
\BIBentrySTDinterwordspacing

\bibitem{Brittain2007ABrushes}
\BIBentryALTinterwordspacing
W.~J. Brittain and S.~Minko, ``{A structural definition of polymer brushes},'' \emph{Journal of Polymer Science Part A: Polymer Chemistry}, vol.~45, no.~16, pp. 3505--3512, 2007.   \url{https://dx.doi.org/10.1002/POLA.22180}
\BIBentrySTDinterwordspacing

\bibitem{Gresham2023AdvancesSurfaces}
\BIBentryALTinterwordspacing
I.~J. Gresham and C.~Neto, ``{Advances and challenges in slippery covalently-attached liquid surfaces},'' \emph{Advances in Colloid and Interface Science}, vol. 315, p. 102906, 2023.   \url{https://dx.doi.org/10.1016/J.CIS.2023.102906}
\BIBentrySTDinterwordspacing

\bibitem{Krasovitski2005DropsPlate}
\BIBentryALTinterwordspacing
B.~Krasovitski and A.~Marmur, ``{Drops down the hill: Theoretical study of limiting contact angles and the hysteresis range on a tilted plate},'' \emph{Langmuir}, vol.~21, no.~9, pp. 3881--3885, 2005.   \url{https://dx.doi.org/10.1021/LA0474565}
\BIBentrySTDinterwordspacing

\bibitem{Orell1971FormationBelow}
\BIBentryALTinterwordspacing
A.~Orell and S.~G. Bankoff, ``{Formation of a dry spot in a horizontal liquid film heated from below},'' \emph{International Journal of Heat and Mass Transfer}, vol.~14, no.~11, pp. 1835--1842, 1971.   \url{https://dx.doi.org/10.1016/0017-9310(71)90050-0}
\BIBentrySTDinterwordspacing

\bibitem{Schwartz2001DewettingFilm}
\BIBentryALTinterwordspacing
L.~W. Schwartz, R.~V. Roy, R.~R. Eley, and S.~Petrash, ``{Dewetting Patterns in a Drying Liquid Film},'' \emph{Journal of Colloid and Interface Science}, vol. 234, no.~2, pp. 363--374, 2001.   \url{https://dx.doi.org/10.1006/JCIS.2000.7312}
\BIBentrySTDinterwordspacing

\bibitem{Shirtcliffe2004ThePrototyping}
\BIBentryALTinterwordspacing
N.~J. Shirtcliffe, S.~Aqil, C.~Evans, G.~McHale, M.~I. Newton, C.~C. Perry, and P.~Roach, ``{The use of high aspect ratio photoresist (SU-8) for super-hydrophobic pattern prototyping},'' \emph{Journal of Micromechanics and Microengineering}, vol.~14, no.~10, p. 1384, 2004.   \url{https://dx.doi.org/10.1088/0960-1317/14/10/013}
\BIBentrySTDinterwordspacing

\end{thebibliography}

\begin{thebibliography}{1}
\providecommand{\url}[1]{#1}
\csname url@samestyle\endcsname
\providecommand{\newblock}{\relax}
\providecommand{\bibinfo}[2]{#2}
\providecommand{\BIBentrySTDinterwordspacing}{\spaceskip=0pt\relax}
\providecommand{\BIBentryALTinterwordstretchfactor}{4}
\providecommand{\BIBentryALTinterwordspacing}{\spaceskip=\fontdimen2\font plus
\BIBentryALTinterwordstretchfactor\fontdimen3\font minus \fontdimen4\font\relax}
\providecommand{\BIBforeignlanguage}[2]{{%
\expandafter\ifx\csname l@#1\endcsname\relax
\typeout{** WARNING: IEEEtran.bst: No hyphenation pattern has been}%
\typeout{** loaded for the language `#1'. Using the pattern for}%
\typeout{** the default language instead.}%
\else
\language=\csname l@#1\endcsname
\fi
#2}}
\providecommand{\BIBdecl}{\relax}
\BIBdecl

\bibitem{Bico2002WettingSurfaces}
\BIBentryALTinterwordspacing
J.~Bico, U.~Thiele, and D.~Qu{\'{e}}r{\'{e}}, ``{Wetting of textured surfaces},'' \emph{Colloids and Surfaces A: Physicochemical and Engineering Aspects}, vol. 206, no. 1-3, pp. 41--46, 2002. \url{https://dx.doi.org/10.1016/S0927-7757(02)00061-4}
\BIBentrySTDinterwordspacing

\bibitem{AlBalushi2022BinarySurfaces}
\BIBentryALTinterwordspacing
K.~M. Al~Balushi, K.~Sefiane, and D.~Orejon, ``{Binary mixture droplet wetting on micro-structure decorated surfaces},'' \emph{Journal of Colloid and Interface Science}, vol. 612, pp. 792--805, 2022. \url{https://dx.doi.org/10.1016/J.JCIS.2021.12.171}
\BIBentrySTDinterwordspacing

\bibitem{Li2017MonostableMaterials}
\BIBentryALTinterwordspacing
Y.~Li, D.~Qu{\'{e}}r{\'{e}}, C.~Lv, and Q.~Zheng, ``{Monostable superrepellent materials},'' \emph{Proceedings of the National Academy of Sciences of the United States of America}, vol. 114, no.~13, pp. 3387--3392, 2017. \url{https://dx.doi.org/10.1073/PNAS.1614667114}
\BIBentrySTDinterwordspacing

\end{thebibliography}


\end{bibunit}
\clearpage

\renewcommand{\thefigure}{S\arabic{figure}}
\renewcommand{\thetable}{S\arabic{table}}

\setcounter{figure}{0}
\setcounter{section}{0}
\setcounter{equation}{0}


  \begin{titlepage}
    \drop=0.2\textheight
    \centering
    \vspace*{\baselineskip}
    \vspace*{100pt}
    \rule{\textwidth}{1.6pt}\vspace*{-\baselineskip}\vspace*{2pt}
    \rule{\textwidth}{0.4pt}\\[\baselineskip]
      
    \LARGE The interplay between topography and contact line pinning mechanisms on flat and superhydrophobic surfaces
    \rule{\textwidth}{0.4pt}\vspace*{-\baselineskip}\vspace{3.2pt}
    \rule{\textwidth}{1.6pt}\\[\baselineskip]
\Large
 \textbf{Supplementary Information}
   
    \vspace*{2\baselineskip}
    \large Mahya Meyari$^{1*}$, Camelia Dunare$^{2}$, Khellil Sefiane$^{3}$, Simon Titmuss$^{1}$, Job H. J. Thijssen$^{1*}$
    \\  \vspace*{1\baselineskip}
    \small{$^{1}$  SUPA School of Physics and Astronomy, The University of Edinburgh, Edinburgh EH9 3FD, UK \\}
    $^{2}$  Scottish Microelectronics Centre, School of Engineering, The University of Edinburgh,  Edinburgh EH9 3FF, UK\\
    $^{3}$  Institute for Multiscale Thermofluids, School of Engineering, The University of Edinburgh, Edinburgh EH9 3FD, UK\\

    \textbf{$^{*}$E-mail:  }\href{mailto:mahya.m@ed.ac.uk}{mahya.m@ed.ac.uk} 
   \textbf{|}  \href{mailto:j.h.j.thijssen@ed.ac.uk}{j.h.j.thijssen@ed.ac.uk}

    \vspace*{2\baselineskip}
   
\vspace*{1\baselineskip}

    \vfill
    
    {\scshape \Large October 2024} \\
    {\large}\par
\end{titlepage}
\begin{bibunit}[IEEEtran]
\section{Surface design}
\begin{figure}[h!]

  \begin{center}
  
  \vspace{-0.1 cm}
    \includegraphics[width=0.4\textwidth]{ 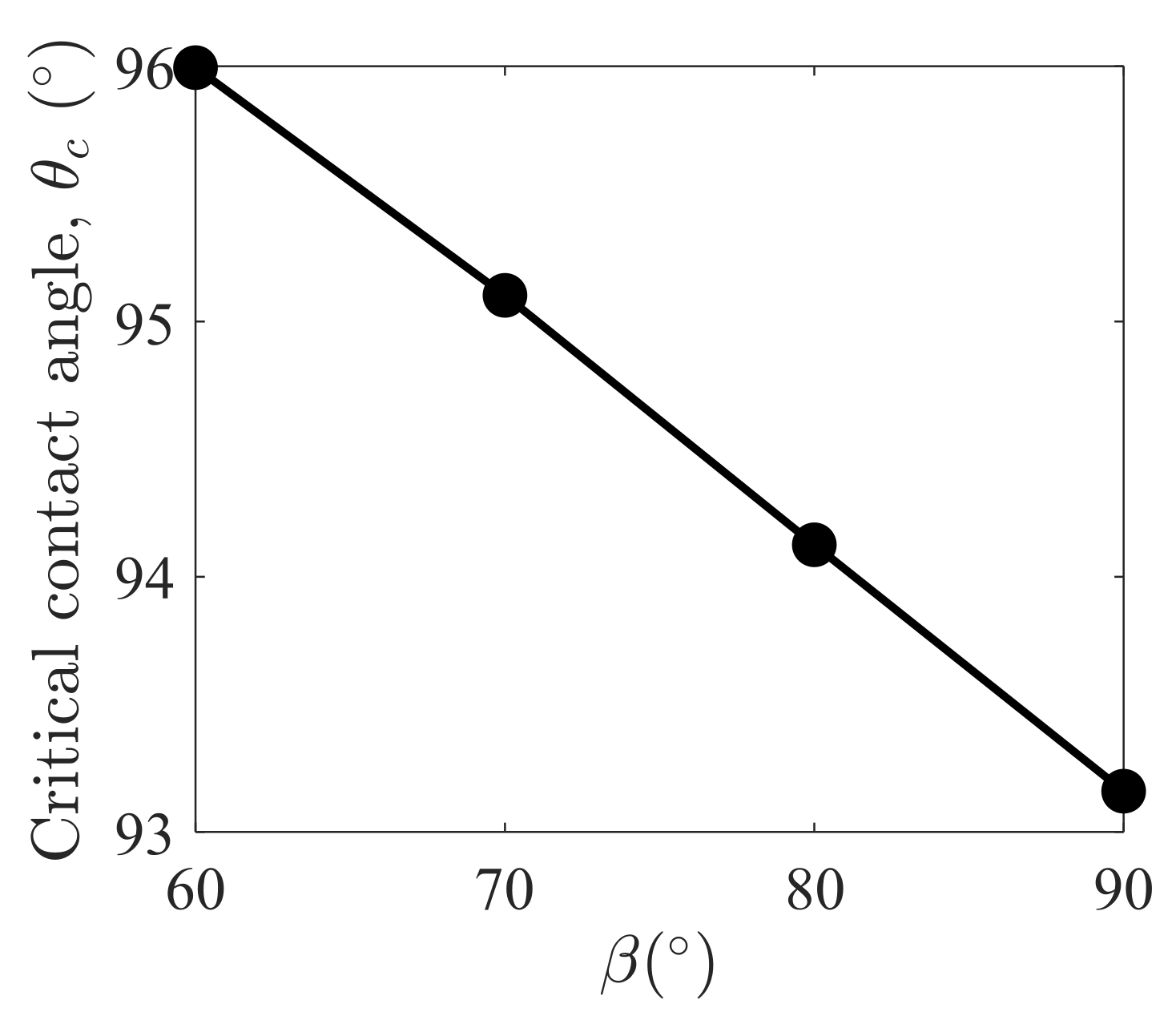}
    \vspace{-0.65 cm}
  \end{center}
\caption{Critical contact angle for the suspended-to-penetrated transition as a function of $\beta$. The values are calculated for $c = 15 \; \mu$m, $l_{v} = l_{h} = 5 \; \mu$m and pillar height $H \approx 50 \; \mu$m.}
  \label{fig-theta_C}
\end{figure}

\begin{figure}[h!]
  \begin{center}
  
  \vspace{-0.28 cm}
    \includegraphics[width=0.8\textwidth]{ 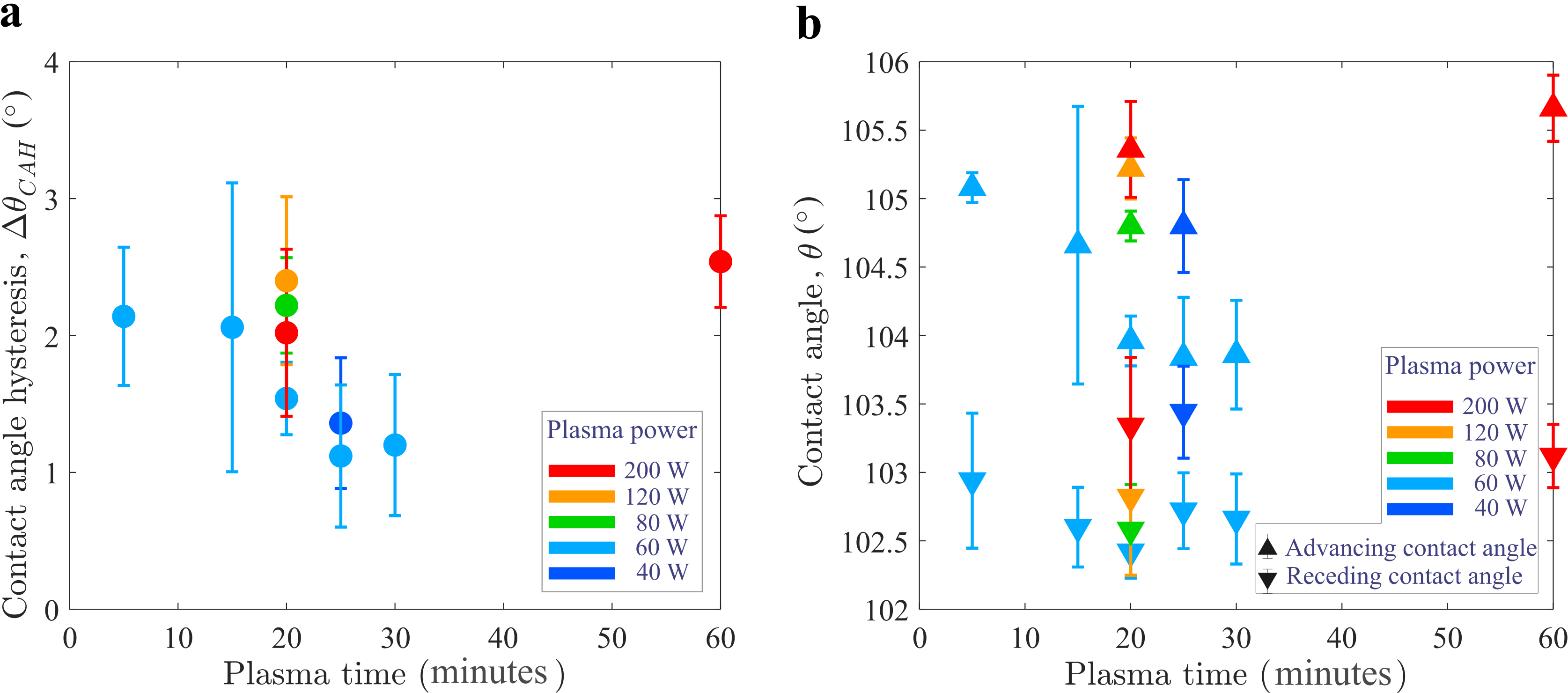}
    \vspace{-0.65 cm}
  \end{center}
\caption{Effect of oxygen plasma time and power on the static wetting properties of the coating \textbf{a)} change of the contact angle hysteresis with plasma time and power; \textbf{b)} change of the advancing and receding contact angles with plasma time and power. The measurements given in this figure were taken by inflating and deflating the water droplet (i.e. volume change method). Error bars present the sample standard deviation for five independent measurements.}
  \label{plasma}
\end{figure}

We have considered a range of pillar geometries from rectangular to triangular shape.
Considering a triangular array design, the pattern can be described by setting the values of \textit{c}, $\beta$, $l_{v}$ and $l_{h}$ as shown in Figure \ref{pillar}. The solid surface fraction and roughness factor can then be calculated as follows:

\begin{figure}[h!]

  \begin{center}
  
  \vspace{-0.1 cm}
    \includegraphics[width=0.5\textwidth]{ 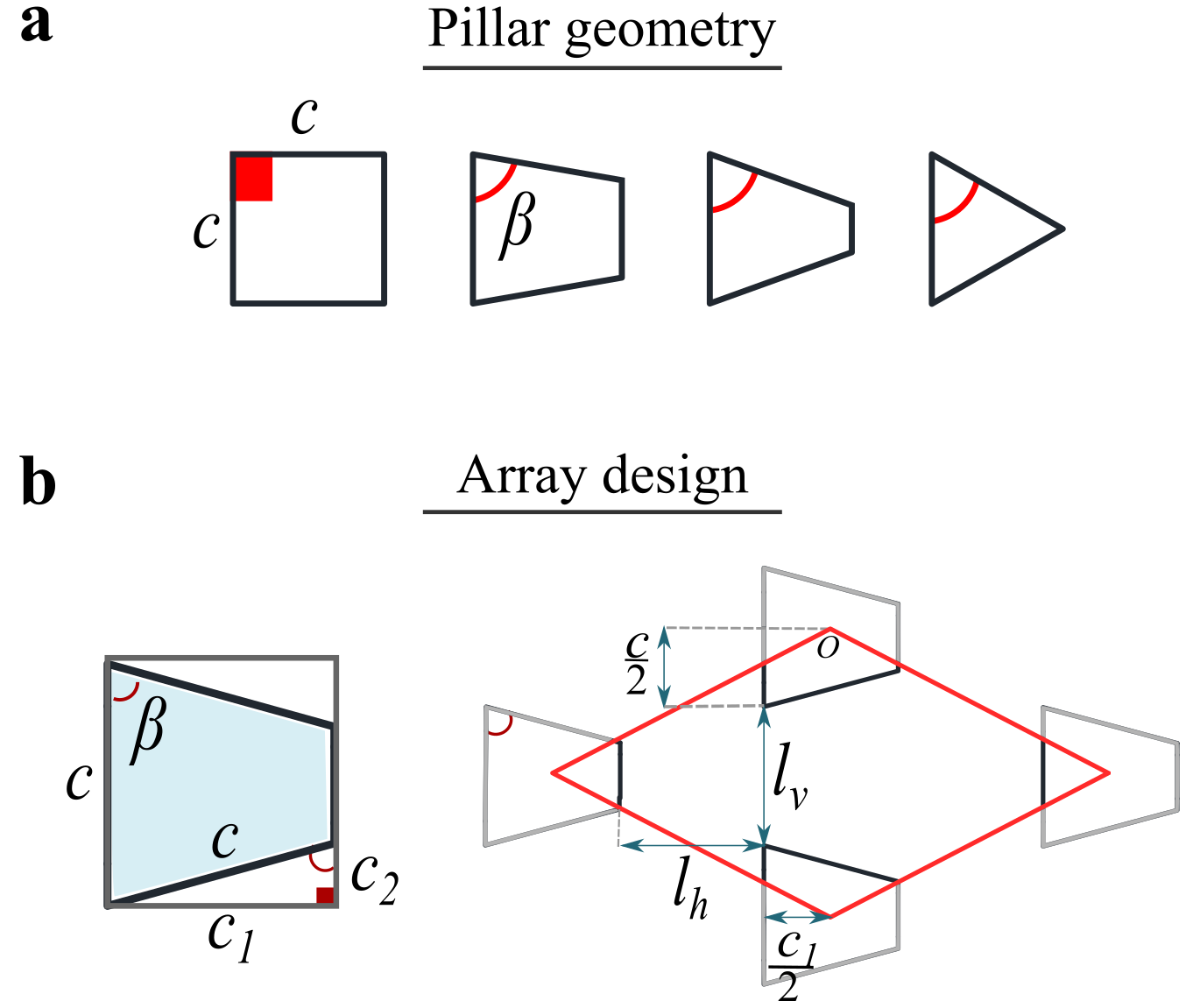}
    \vspace{-0.65 cm}
  \end{center}
\caption{Design of the patterns: \textbf{a)} \textit{c} and $\beta$ are the parameters that define the pillar geometry. \textit{c} remains constant for all the pillars; \textbf{b)} the triangular array is described by the vertical and horizontal gap between the pillars ($l_{v}$ and $l_{h}$ respectively). We designed four pillar geometries with $\beta = 60^{\circ}, \;70^{\circ}, \;80^{\circ}$ and $90^{\circ}$. We set $c = 15 \; \mu$m, $l_{v} = l_{h} = 5 \; \mu$m and pillar height $H \approx 50 \; \mu$m in all the arrays.}
  \label{pillar}
\end{figure}
\begin{itemize}
    \item {Calculating the solid surface fraction ($\phi_{s}$):} 
    
Solid surface area ($S_{s}$):
\begin{equation} \label{pillarDesign}
S_{s} = cc_{1}-2\frac{c_{1}c_{2}}{2}
\end{equation}
where $c_{1} = c \sin \beta$  and $c_{2} = c \cos \beta$.

Total surface area ($S_{t}$):
\begin{equation} \label{S_t}
S_{t} = \frac{(l_{v}+c)(2l_{h}+2c_{1})}{2} = (l_{v}+c)(l_{h}+c\sin\beta)
\end{equation}

Therefore, the solid surface fraction can be calculated as:
\begin{equation} \label{phi_s}
\phi_{s} = \frac{c\sin\beta(c-c\cos\beta)}{(l_{v}+c)(l_{h}+c\sin\beta)}
\end{equation}

\item Calculating the Wenzel roughness factor ($r_{W}$):
To calculate the total area of possible contact between the liquid and solid, we need to calculate the area of the vertical walls, and consequently, the perimeter of the pillars, \textit{p}:
\begin{equation} \label{Pw}
p = 4c - 2c_{2} = 2c + 2c (1-\cos\beta)
\end{equation}
For a pillar height of \textit{H}, the roughness factor ($r_{W}$) is calculated as:
\begin{equation} \label{rw}
r_{W} = \frac{S_{t}+S_{\rm{vertical-walls}}}{S_{t}} = \frac{S_{t}+Hp}{S_{t}} = \frac{(l_{v}+c)(l_{h}+c\sin\beta) + H(2c + 2c (1-\cos\beta))}{(l_{v}+c)(l_{h}+c\sin\beta)}
\end{equation}
\item Calculating the critical contact angle for the suspended-to-penetrated transition ($\theta_{c}$):
Using the surface energy argument, the following equation can be used to estimate the smallest contact angle on the flat surface for which the suspended state would be more favourable than the penetrated state on the structured surface \cite{Bico2002WettingSurfaces}:

\begin{equation} \label{theta_c}
\theta_{c}=\cos^{-1}(-\frac{1-\phi_{s}}{r_{W}-\phi_{s}})
\end{equation}
\end{itemize} 
 The calculated values of $\theta_{c}$ for the designed surfaces are given in Figure \ref{fig-theta_C}. On a QLS coating, the water contact angle is normally larger than $100^{\circ}$. Since we are interested in suspended droplets, $\theta_{c}$ should be smaller than the intrinsic contact angle on the coating, which was the case in our designed surfaces. The experimental observations (e.g. droplets remaining spherical \cite{AlBalushi2022BinarySurfaces} and rolling off the tilted surface) were consistent with this theoretical prediction - indicating the water droplets were suspended on the micropillars. It is also worth mentioning that since even the intrinsic receding contact angle is larger than $\theta_{c}$ as calculated above (see Section 3.2 and Figure \ref{plasma} for the contact angle values), the designs meet the surface energy criterion for the monostability of the suspended state \cite{Li2017MonostableMaterials}.

\section{QLS coating optimisation}

\section{Surface characterisation}

\begin{figure}[h!]
  \begin{center}
 
  \vspace{-0.28 cm}
    \includegraphics[width=1\textwidth]{ 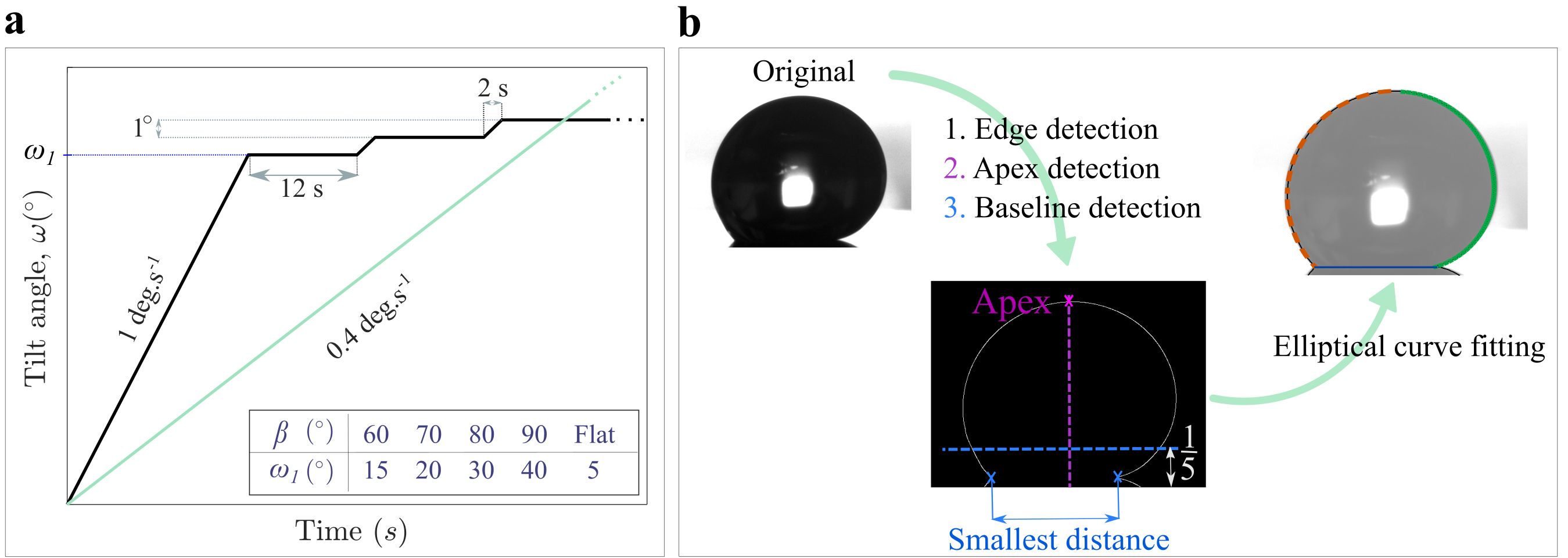}
    \vspace{-0.65 cm}
  \end{center}
\caption{\textbf{a)} Stage tilting program for the tilting-method measurements given in the main text. The black curve was used to obtain the data in Figure 3d. The green curve shows the program used to collect the data in the Figure 4 (adaptation). \textbf{b)} Image analysis for the tilting method experiments with baseline detection. Canny algorithm is used for edge detection, followed by separating the detected shape into two parts from the apex and baseline detection. Two elliptical curves are then fitted to the left and right parts of the droplet. The camera was tilted with the stage; therefore, the baseline remained horizontal.}
  \label{tiltCurve}
\end{figure}

\end{bibunit}

\end{document}